\documentclass[%
 reprint,
superscriptaddress,
longbibliography,
 amsmath,amssymb,
 aps,
 prl,
floatfix,
]{revtex4-1}

\usepackage[mode=buildnew]{standalone}
\usepackage{tikz}
\usepackage{bm}
\usepackage{physics}
\usepackage{xcolor}
\usepackage{mathrsfs}

\usepackage{graphicx}
\usepackage{dcolumn}
\usepackage{bm}
\usepackage[normalem]{ulem}
\usepackage{subfig}

\usepackage{graphicx}
\usepackage{bm}

\newcommand{\<}{\langle}
\renewcommand{\>}{\rangle}

\newcommand{\comments}[1]{}
\newcommand{\ba}{\begin{align}}
\newcommand{\ea}{\end{align}}

\newcommand{\trR}[1]{\text{Tr}_{\scriptscriptstyle R}\big[#1\big]}

\newcommand{\id}{\mathbb{I}}

\begin{document}

\title{Quantum work statistics at strong reservoir coupling}
\author{Owen Diba}
\author{Harry J.~D. Miller}
\author{Jake Iles-Smith}
\author{Ahsan Nazir}
\affiliation{Department of Physics and Astronomy, The University of Manchester, Manchester M13 9PL, UK}

\date{\today}
\begin{abstract}
Determining the statistics of work done on a quantum system while strongly coupled to a reservoir is a formidable task,
requiring the calculation of the full eigenspectrum of the combined system and reservoir. Here we show that this issue can be circumvented by using a polaron transformation that maps the system into a new frame where weak-coupling theory can be applied. 
Crucially, this polaron approach reproduces the Jarzynski fluctuation theorem, thus ensuring consistency with the laws of stochastic thermodynamics. We apply our formalism to a system driven across the Landau-Zener transition, where we identify clear signatures in the work distribution arising from a non-negligible coupling to the environment. Our results provide a new method for studying the stochastic thermodynamics of driven quantum systems beyond Markovian, weak-coupling regimes.
\end{abstract}

\maketitle

In standard approaches to thermodynamics it is generally assumed that interactions between a system and its environment play a negligible role. However, rapid developments in experimental control of mesoscopic and quantum systems \cite{Alkauskas_2014,Wei14,Bosman17,potovcnik2018studying,Brash19,Maier2019,Clear20,Takahashi2020,Bando22,Fischer23} have generated significant interest in exploring how non-negligible interactions and structured environments can modify both the equilibrium and non-equilibrium behaviour of strongly coupled systems \cite{Ingold2008,Strasberg2016a,perarnau-llobet2018,Hsiang2018,gelbwaser-klimovsky2015,Strasberg2019b,Rivas2020,Talkner2020}.


The first key departure from weak-coupling thermodynamics lies in the fact that the equilibrium state of the system generally differs from the familiar Gibbs state. Studies of both exactly-solvable and approximate models of strongly coupled quantum systems provide evidence that, when left to equilibriate, the system reaches the partial trace of a global Gibbs state with respect to both system and environment \cite{Grabert1984,Subasi2012,Thingna2012,Cresser2021a,Trushechkin2022}. Such deviations from standard equilibrium have also lead to recent extensions of stochastic thermodynamics to the strong coupling regime out of equilibrium. For example, one can derive an extension of the work fluctuation theorems when a system is driven away from a non-Gibbsian equilibrium state \cite{Campisi2009a}. However, one issue with these formal approaches is that the work distribution $P(W)$ is formidably difficult to compute in strong-coupling regimes since it is obtained from projecting the total system and environment into global energy eigenstates at the start and end of the process \cite{Campisi2009a,Talkner2009,Talkner2020}.  Ultimately this means that, with the exception of relatively few exactly solvable models \cite{campisi2009b,PhysRevB.91.224303,Aurell2017a,funo2018}, methods for calculating $P(W)$ in open systems outside the weak coupling regime still remain limited. This is a significant issue for accurately modelling thermodynamic behaviour in a wide range of quantum system where we expect the interaction energy to contribute significantly to the work done on the system. The inadequacy of weak coupling approaches to quantum dynamics has been experimentally demonstrated in diverse
platforms such as quantum dots~\cite{Wei14,Brash19}, defect centres in materials~\cite{Alkauskas_2014,Fischer23}, superconducting circuits~\cite{Bosman17, Bando22}, and organic molecules~\cite{Clear20}. The strong coupling regime is therefore of key importance to quantum thermodynamics, and obtaining accurate predictions for work statistics in microscopic systems  inevitably requires more sophisticated methods that go beyond weak coupling approximations.

One significant issue is that numerically exact techniques for simulating open system dynamics, such as discrete time path-integral methods~\cite{makri1995tensor,strathearn2018efficient,cygorek2022simulation}, are formulated to describe the dynamics of the reduced state of the system, and therefore do not have straightforward access to the eigenspectrum of the composite system and environment required to construct the full work distribution. In this paper we introduce an alternative approach based on the polaron framework~\cite{nazir2009correlation,jang2008,mccutcheon2011} that is not hampered by these issues.
The polaron method
uses a unitary transformation to dress system states with vibrational modes of the environment, capturing the dominant non-Markovian effects and allowing the system-environment interaction to be included  
non-perturbatively. The polaron framework has been successfully adapted to study resonant energy transfer in photosynthetic complexes \cite{kolli2011electronic,jang2011}, exciton-phonon interactions in semiconductor quantum dots \cite{mccutcheon2010quantum}, and heat exchange statistics in the 
spin-boson model with multiple reservoirs \cite{wang2015,maria2021}. Within the polaron transformed frame, we are able to derive a generalised master equation (GME) that governs the evolution of the characteristic function associated with $P(W)$ beyond weak coupling regimes. This can be used to infer the full statistics of work and reproduces the Jarzynski fluctuation theorem \cite{Campisi2009a} as required for thermodynamic consistency. To demonstrate our approach, we consider a Landau-Zener (LZ) driving protocol on the system Hamiltonian. We find that stronger coupling creates new peaks in the distribution as well as imparting significant broadening and renormalisation, indicating clear signatures in the work statistics arising from interactions with the reservoir that are either not captured or severely underestimated by a weak-coupling approach.

\emph{Generalised Master Equation for Quantum Work--}
We begin by considering the spin-boson Hamiltonian, 
  $  H(t)=H_S(t)+H_R+\sigma_z \otimes \sum_k g_k (b_k^\dagger +b_k), $
where the time-dependent system Hamiltonian is given by
    $H_S(t)=(\omega_0(t)\sigma_z+\Delta \sigma_x)/2.$
Here $\omega_0(t)$ is a time-varying energy spacing and $\Delta$ is a fixed tunneling coefficient. 
The reservoir Hamiltonian is given by $H_R=\sum_{k} \omega_k b_k^\dagger b_k$, with $b_k$ the annihilation operator for the $k$th mode of frequency $\omega_k$. The final term in $H(t)$ describes the system-reservoir interaction with coupling constants $g_k$. 
This interaction is fully characterised by a super-Ohmic spectral density $J(\omega)=\alpha\omega_c^{-2} \omega^3\exp(-\omega/\omega_c)$, 
with cutoff frequency $\omega_c$ and interactions strength $\alpha$.  
The composite state at time $t$ is given by 
$\rho(t)=U(t)\rho(t_i) U^\dagger(t)$ with initial state $\rho(t_i)$ and time-ordered unitary $U(t)=\overleftarrow{\mathcal{T}}\text{exp}\big(-i\int^{t}_{t_i} dt' H(t')\big)$. 

We now wish to compute the statistics of work done on the system along a driving protocol with duration $t_f-t_i$, that is we are interested in the total energetic change of the system and environment during the protocol, including any stochastic fluctuations of these quantities. 
Independent of the coupling strength, the work done can be defined using the two-point measurement protocol, where a projective energy measurement of the global Hamiltonian $H(t)$ 
is made at the initial and final times of the protocol, with the work done $W$ defined as the resulting energy difference~\cite{Campisi2009a}. 
The resulting distribution $P(W)$ can be characterised by its characteristic function (CF), 
\begin{align}\label{eq:CF}
    \Phi(\eta) =\int^\infty_{-\infty}e^{i\eta W}P(W)dW=\tr{K(t_f,\eta)}.
\end{align}
Here we define the \textit{work characteristic operator} (WCO) \cite{Talkner2007b}, 
\begin{align}\label{eq:WCOin}
    K(t_f,\eta):=e^{i \eta H(t_f)}U(t_f)e^{-i\eta H(t_i)}\bar{\rho}(t_i) U^\dagger(t_f),
\end{align}
where $\bar{\rho}(t_i)=\sum_n \Pi^i_n \rho(t_i) \Pi^i_n$ is the initial state dephased in the energy eigenbasis of $H(t_i)$, with $\Pi^i_n$ a projector onto the $n$th eigenstate.
Without further approximations, computing the CF even numerically is challenging due to the presence of the interaction. One immediate solution would be to assume the interaction is sufficiently weak, allowing one to derive a Lindblad-like master equation for the WCO~\eqref{eq:WCOin} \cite{esposito2009,silaev2014,suomela2015,liu2016}. However, our aim is to probe regimes where this approximation is not appropriate, and so we require an alternative method for computing $K(t,\eta)$. 
Our approach is to apply a trace-preserving operation to the WCO such as 
$
    K(t_f,\eta)\mapsto K_P(t_f,\eta)=e^{P} K(t_f,\eta) e^{-P}.
$
This leaves the CF invariant so that $\Phi(\eta)=\tr{K_P(t_f,\eta)}$. For our spin-boson model we choose the unitary polaron transformation \cite{mccutcheon2011}, where $P=\sigma_z\otimes\sum_k (g_k/\omega_k) (b_k^\dagger-b_k)$. 
This transformation redraws the boundary between the system and environment by dressing the system states with vibrational modes. 
These dressed system states account for a significant part of the interaction,
and the residual part of the interaction in this frame can be treated perturbatively. 
The transformed WCO is given by
 $  K_P(t_f,\eta)=e^{i \eta H_P(t_f)}U_P(t_f)e^{-i\eta H_P(0)}\bar{\rho}_P(t_i) U_P^\dagger(t_f),
$
where $\bar{\rho}_P(t_i)=e^P \bar{\rho}(t_i) e^{-P}$ and $U_P(t)=\overleftarrow{\mathcal{T}}\text{exp}\big(-i\int^{t_f}_{t_i} dt' H_P(t')\big)$ is the evolution operator with respect to the polaron-transformed Hamiltonian
$
    H_P(t):=e^{P} H(t)e^{-P}=H_{PS}(t)+H_R+V_{P}.
$ 
Here we identify the polaron system Hamiltonian
\begin{align}\label{eq:ham_PS}
    H_{PS}(t)=\frac{1}{2}\omega_0(t) \sigma_z + \frac{1}{2}\Delta \kappa \sigma_x,
\end{align}
where $0\leq \kappa\leq 1$ denotes the \textit{polaron renormalisation constant} that lowers the tunneling coefficient of our transformed system. The degree of renormalisation is dependent on the structure of the reservoir,  and can be computed from its spectral density according to
\begin{align}\label{eq:kappa}
    \kappa:=\exp\bigg(-2 \int^\infty_0 d\omega \frac{J(\omega)}{\omega^2}\coth \big(\beta\omega/2\big)\bigg). 
\end{align}
Furthermore, we have identified a new interaction term in the polaron frame given by
\begin{align}\label{eq:interaction}
     V_P=\frac{\Delta}{2}\bigg(\sigma_x\otimes\xi_x+\sigma_y\otimes\xi_y\bigg), 
\end{align}
where $\xi_x=\frac{1}{2}(\xi_+ +\xi_{-})-\kappa$ and $\xi_y=\frac{i}{2}(\xi_+ -\xi_{-})$, with $\xi_{\pm}=\Pi_k D(\pm\alpha_k)$. This new interaction can now be treated in a perturbative manner quantified by the parameter \cite{mccutcheon2011}
\begin{align}\label{eq:interaction_g}
    g:=\frac{\Delta}{2}\sqrt{(1+\kappa^4)/2}.
\end{align}
Taking a weak-coupling approximation in this new frame will allow us to determine the work statistics of driven systems that are strongly coupled to the reservoir with respect to the original frame. Note that we have assumed a cubic spectral density in~\eqref{eq:kappa}, which is applicable e.g.~to solid-state
systems coupled to acoustic phonons \cite{nazir2016}. However, our methods can be straightforwardly extended beyond super-Ohmic spectral densities using variational approaches \cite{harris1985}.

Let us assume that the system and environment have initially thermalised to a global Gibbs state with respect to the Hamiltonian in the original frame, $\rho(t_i)=\pi^{eq}(t_i):=e^{-\beta H(t_i)}/\tr{e^{-\beta H(t_i)}}$
with $\beta=1/k_B T$ an inverse temperature. As our first approximation we suppose that the interaction strength in the polaron frame is weak enough relative to the reservoir temperature so that $\beta^2 g^2\ll 1$. We can then neglect any correlations in the initial state and contributions of the interaction to the initial and final energy measurements, so that the polaron transformed WCO simplifies to  
\begin{align}\label{eq:WCO2}
    \nonumber K_P(t_f,\eta)\simeq &e^{i \eta (H_{PS}(t_f)+H_B)}U_P(t_f)e^{-i \eta (H_{PS}(t_i)+H_B)} \\
    & \ \ \ \ \ \ \ \ \ \ \ \ \ \times \big(\pi^{eq}_{PS}(t_i)\otimes \pi^{eq}_R\big) U_P^\dagger(t_f),
\end{align}
where $\pi^{eq}_R$ is a Gibbs state of the bare reservoir with Hamiltonian $H_R$ while $\pi^{eq}_{PS}(t_i)$ is a Gibbs state of the polaron system with the initial Hamiltonian $H_{PS}(t_i)$. We emphasise that this approximation is valid in strong coupling regimes, unlike the typical weak coupling limit in which $\rho(t_i)\simeq \pi^{eq}_{S}(t_i)\otimes \pi^{eq}_R$ \cite{Talkner2009}.

\begin{figure*}[t!]
    \centering
    \includegraphics[width=\textwidth]{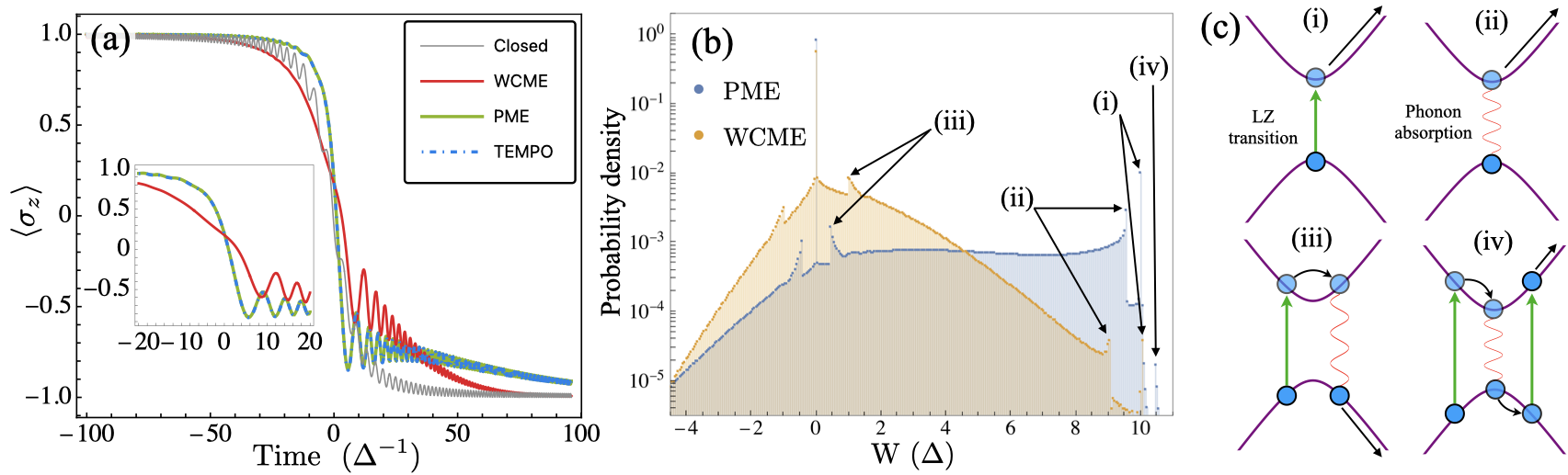}
    \caption{(a) Population dynamics of the dissipative Landau-Zener model predicted by different dynamical models; closed unitary evolution, the weak-coupling master equation (WCME), the polaron master equation (PME), and an exact numerical simulation provided by TEMPO~\cite{strathearn2018efficient}. The insets shows the dynamics near to the avoided crossing.
    (b) Probability distribution of the dissipative LZ model as given by polaron (PME) and weak-coupling (WCME) generalised master equations. The CF step width is  $\delta\eta=0.05$, the maximum counting-field value is $\eta_\text{max}=500$, and the bin width is $\delta W=0.05\Delta$.
     (c) Figures (i-iv) show the mechanisms leading to the labelled peaks in the probability distribution.
    Other parameters used in (a) and (b) are $\nu=0.1\Delta^2$, $t_0=-100/\Delta$, $t_f=100/\Delta$, $\alpha=0.4$, $\beta=1/\Delta$, $\omega_c=10\Delta$. }
     \label{fig:hist}
\end{figure*}

To obtain the work statistics we derive a master equation for the WCO of the system degrees of freedom in the polaron frame, which is obtained by taking a partial trace over the reservoir so that $K_{PS}(t_f,\eta)=\tr_\mathrm{R}\left({K_{P}(t_f,\eta)}\right)$. The derivation of an equation for the evolution of $K_{PS}(t_f,\eta)$ generalises the projection operator technique  
used to obtain an adiabatic master equation for the polaron state itself, and details are provided in the Supplementary Material. This rests on a number of assumptions such as $(i)$ the Born-Markov approximation, stating that the coupling in the polaron frame is much smaller than the inverse of the decay time of the reservoir correlation function and $(ii)$ the adiabatic approximation, which ensures that the system Hamiltonian does not change too rapidly in time relative to its energy gaps and bath correlation time. Note that applying the Born-Markov approximation in the polaron frame is {\emph{not}} equivalent to applying it in the original frame, and retains the dominant non-Markovian effects in the dynamics. Under these assumptions the general form of equation for the polaron WCO is then
\begin{align}\label{eq:lind}
    \partial_t K_{PS}(t,\eta)=\mathcal{L}_{t}[K_{PS}(t,\eta)]+\mathcal{W}(t,\eta)[K_{PS}(t,\eta)],
\end{align}
where we define the work superoperator
\begin{align}\label{eq:WCO_pol}
\mathcal{W}(t,\eta)(.)=\big[\partial_t e^{i\eta H_{PS}(t)}\big]e^{-i\eta H_{PS}(t)}(.),
\end{align}
and $\mathcal{L}_{t}(.)=-i[H_{PS}(t)+H_{PLS}(t),(.)]+ \mathcal{D}_t(.)$ is an adiabatic Lindblad generator with $H_{PLS}(t)$ the polaron lamb shift and $\mathcal{D}_t(.)$ the dissipator, which are both derived and defined in the Supplementary Material.
Physically, the dissipator drives transitions between the instantaneous energy eigenstates of the polaron system, defined as $\ket{\varepsilon_+(t)}=\cos(\theta_t/2)\ket{1} + \sin(\theta_t/2)\ket{0}$ and $\ket{\varepsilon_-(t)}=-\sin(\theta_t/2) \ket{1}+ \cos(\theta_t/2) \ket{0}$ with $\theta_t = \arctan(\Delta \kappa/\omega_{0}(t))$ and energy eigenvalues $\varepsilon_\pm(t) = \pm\frac12\omega(t)$, where $\omega(t) = \sqrt{\omega_0(t)^2+\kappa^2 \Delta^2}$ is the transition frequency. This contrasts with the standard weak-coupling master equation (WCME) for time-dependent driving \cite{albash2012}, whose dissipator causes transitions between eigenstates of the original system Hamiltonian $H_S(t)$ (see Supplementary Material).
In Fig.~\ref{fig:hist}(a) we have benchmarked the population dynamics predicted by the polaron master equation (PME) against the numerically exact TEMPO method~\cite{strathearn2018efficient}. The PME faithfully describes the system dynamics across all timescales. Further benchmarks can be found in the Supplementary Material.

A key property of the generator $\mathcal{L}_{t}$ is that it has an instantaneous thermal fixed point with respect to the polaron Hamiltonian, so that $\mathcal{L}_{t}[\pi^{eq}_{PS}(t)]=0$. When this detailed balance condition is combined with the equation of motion for WCO~\eqref{eq:lind}, then  a standard procedure \cite{chetrite2012quantum} can be used to obtain the Jarzynski fluctuation relation:
\begin{align}
     \langle e^{-\beta W}\rangle=e^{-\beta \Delta F_{PS}}; \ \ \ 
\end{align}
where $\Delta F_{PS}$ is the change in the polaron free energy, $F_{PS}(t)=-k_B T \ln (2\cosh[\frac{\beta}{2}\omega(t)])$. In fact, we recognise that this result is equivalent to the exact fluctuation theorem derived in \cite{Campisi2009a} since the polaron's free energy is approximately equal to the original system's mean-force free energy $F_S^*(t)=F_{SR}(t)-F_R$ \cite{Jarzynski2004}. The latter is defined as the difference between the total free energy $F_{SR}(t)$ and that of the bare reservoir $F_R$ \cite{Ford1985,Jarzynski2004, Gelin2009a,Campisi2009a,Philbin2016,miller2018,Jarzynski2016}, and the approximation $F_{PS}(t)\simeq F_{S}^*(t)$ holds under our previous assumption $\beta^2 g^2\ll 1$. One can therefore conclude that the work distribution predicted by our equation~\eqref{eq:lind} is consistent with the laws of stochastic thermodynamics \cite{Talkner2020}.

\emph{Work Statistics---}~
To illustrate our polaron approach we shall consider the specific case of a two level system driven over a Landau-Zener crossing. In this case, the time dependent driving in the Hamiltonian may be written as $\omega_0(t) = \nu t$, where $\nu$ is the rate at which the driving occurs.  In all following calculations, we choose  $\nu=0.1 \Delta^2$ and start and end the protocol far from the avoided crossing $-t_0=t_f=100/\Delta$, ensuring that the adiabatic condition is satisfied. To ensure the validity of the PME we work in the scaling limit $\Delta/\omega_c\ll 1$ \cite{nazir2016,mccutcheon2010quantum}, taking $\omega_c=10\Delta$. Further details about the relevant parameter regimes for the full WCO equation~\eqref{eq:lind} can be found in the Supplementary Material.  To calculate the work distributions we sample the CF via the solution to the generalised PME over a range of counting-field values at intervals $\delta\eta$ up to some cut-off $\eta_{\text{max}}$, before performing a numerical inverse Fourier transformation on this data and integrating over fixed intervals (bins) $\delta W$ to obtain the work probability density function.

In Fig.~\ref{fig:hist} we show polaron and weak-coupling predictions for the quantum work distribution for a choice of $\alpha$ well into the strong coupling regime. The most immediate difference in the open system distribution is that it is continuous in contrast to the discrete distribution of the closed system. Since the bath is composed of a continuum of modes, it may exchange energy with the system at any of the continuous range of transition frequencies that the system moves through during the protocol. In analogy to the closed system, both the polaron and weak-coupling predicted work distributions show peaks at $W=0$ and $W = \Delta E$ corresponding to the adiabatic and non-adiabatic system trajectories (see 
Fig.~\ref{fig:hist}(i) for a schematic of the latter process~\footnote{Both the weak coupling and polaron theory also show a peak at $W=-\Delta E$, however this is heavily suppressed when initialising the system in thermal equilibrium far from the anti-crossing.}).
Interestingly, the likelihood of non-adiabatic transitions is underestimated by over two orders of magnitude in the weak-coupling approach. This is a direct consequence of it ignoring the polaron renormalisation of the energy gap to $\Delta\kappa$, with $\kappa$ given by Eq.~(\ref{eq:kappa}), meaning that the system dynamics must actually be slower to maintain the same degree of adiabaticity.

Unlike a closed system model of a Landau-Zener transition, both the weak-coupling and polaron probability distributions show a rich structure due to phonon-mediated processes. 
The simplest phonon process is shown schematically in Fig.~\ref{fig:hist}(ii), where a phonon excites the system over the avoided crossing, which then continues to evolve adiabatically in the excited state. This requires $W = \Delta E - \Delta\approx 9\omega(t_f)$ of work according to the weak coupling theory, which is increased to $W = \Delta E - \Delta\kappa\approx 9.5\omega(t_f)$ in the polaron approach due to the renormalisation of the avoided crossing.

The peaks in the distribution either side of the origin, which are due to a heat exchange event at the avoided crossing in combination with a nonadiabatic transition (Fig.~\ref{fig:hist}(iii)), can be interpreted similarly and are brought closer together in the polaron predictions due to renormalisation. 
While renormalisation tends to shift such peaks, we also observe that strong coupling tends to increase the likelihood of larger amounts of work being done on the system. 
One explanation for this is routed in the fact that, as mentioned, non-adiabatic transitions are less suppressed in the PME. From a thermodynamic perspective, this leads to a greater probability of excitation and thus 
an increased likelihood 
of dissipation, which here corresponds to larger amounts of work being done to the system. 
As a final observation, the polaron approach predicts another type of process not apparent in the weak coupling theory and given schematically by Fig~\ref{fig:hist}(iv), whereby a LZ transition excites the system, which emits a phonon, before a second LZ transition re-excites the system.
The total work done in this situation is $W = \Delta E + \Delta\kappa\approx10.5\omega(t_f)$.

\begin{figure}
\centering
    \includegraphics[width=\columnwidth]{gridplot.png}
    \caption{Work probability distributions as predicted by (a) the weak coupling (WCME) and (b) the polaron (PME) generalised master equations for increasing system-environment coupling strengths $\alpha$. We use the same driving parameters and temperature as in Fig.~\ref{fig:hist}.}
    \label{fig:alpha_comp}
\end{figure}

These key differences between the polaron 
and weak-coupling predictions 
are highlighted in Fig.~\ref{fig:alpha_comp}, where we compare their respective work distributions 
for increasing $\alpha$. 
The work cost of the system transitioning across the energy gap  
remains constant in the weak coupling model as it incorrectly predicts the gap to be completely independent of the reservoir coupling strength. This is reflected in Fig.~\ref{fig:alpha_comp}(a), where the locations of the various peaks are independent of $\alpha$. In contrast, the polaron work distribution predictions given in Fig.~\ref{fig:alpha_comp}(b) show monotonic 
shifts with $\alpha$ in the locations of peaks involving phonon transitions,   
precisely due to the 
renormalisation of the energy gap. Furthermore, Fig.~\ref{fig:alpha_comp} illustrates that even the qualitative shapes of the work distributions can vary significantly between the polaron and weak-coupling predictions, despite agreeing at very small $\alpha$. 
From these work distributions we can also consider the impact strong reservoir coupling has on the average work done and its fluctuations. 
A more detailed comparison of the mean work and its variance for the polaron and weak-coupling predictions is included in the Supplementary Material, but the main trends can already be seen. In general, for our chosen parameters we find that the weak-coupling theory typically underestimates the average work and its variance, with this discrepancy only becoming negligible at very small couplings. 

\textit{Conclusion---} We have studied the full-counting statistics of work done in the time-dependent spin-boson model beyond the weak coupling regime using the polaron framework. We illustrated our general approach by solving the dissipative Landau-Zener model, where the polaron framework indicated significant increases in both average dissipation and work fluctuations in contrast to predictions from the weak coupling master equation due to renormalisation of system parameters. 
This serves to highlight the limitations of the usual weak-coupling approximation, as even modest increases in the coupling can have a noticeable impact on the stochastic thermodynamics of the system. Our results lay the groundwork for further investigations into how correlations, non-Markovianity and other strong coupling effects can influence the full statistics of work in driven open quantum systems.

\emph{Acknowledgements. } O. D. acknowledges the EPSRC for PhD funding support. H. M. acknowledges funding from a Royal Society Research Fellowship (URF/R1/231394), and the Royal Commission for the Exhibition of 1851. We thank Ali Raza Mirza for helpful discussions.

\bibliography{mybib.bib}

\appendix

\widetext

\section{Deriving the WCO master equation}\label{appA}

\
In this section we provide a derivation of the equation~(12) of the main manuscript, starting from the polaron-transformed work characteristic operator (WCO) in equation (11) of the manuscript. We write the interaction Hamiltonian~(9) in a generic form
\begin{align}
    V_P=\sum_{a} A_a\otimes B_a,
\end{align}
where are $A_a(B_a)$ Hermitian system (bath) operators. The derivation closely follows a projection operator approach used by \cite{liu2018a} to obtain an equation for the WCO with a standard weak-coupling adiabatic master equation. We start by differentiating~11 and find that it satisfies the differential equation 
\begin{multline}\label{eq:}
	\partial_t K_P(t, \eta) = -i\big[H_P(t),K_P(t, \eta)\big]-i \big[e^{i \eta (H_{PS}(t)+H_R)},V_P\big]e^{-i \eta (H_{PS}(t)+H_R)} K_P(t, \eta)+ \left[\partial_t e^{i \eta H_{PS}(t)}\right]e^{-i \eta H_{PS}(t)}K_P(t, \eta).
\end{multline}
with the initial condition $K_P(0)=\sum_{i,j}\Pi_{ij}(0)\rho(0)\Pi_{ij}(0)$, 
where $\rho_P(0)$ is the initial density matrix of the system-bath in the polaron frame and $\Pi_{ij}$ is the projection operator at the initial time that projects onto the eigenspace corresponding to the polaron frame system eigenenergy $\varepsilon_i(0)$ and the bath eigenenergy $\varepsilon^B_j$ i.e. 
\begin{align}
    H_{PS}\Pi_{ij}(0) &= \varepsilon_i(0)\Pi_{ij}(0),\\
    H_{B}\Pi_{ij}(0) &= \varepsilon^B_j(0)\Pi_{ij}(0).
\end{align}
To simplify notation we define a set of superoperators:
\begin{align}
	\mathcal{V}(t)[\cdot] &\equiv -i[H_P(t),(\cdot)],\\
	\mathcal{J}_{\eta}(t)[\cdot]&= -i\big[e^{i \eta (H_{PS}(t)+H_R)},V_P\big]  e^{-i \eta (H_{PS}(t)+H_R)}(\cdot), \\
	\mathcal{W}_{\eta}(t)[\cdot] &= [\partial_t e^{i \eta H_{PS}(t)}]e^{-i \eta H_{PS}(t)}(\cdot),
\end{align}
where the subscript $\eta$ denotes a dependence on the counting parameter. Next we move into to the interaction picture with respect to $H_0(t) = H_{PS}(t)+H_R$ and write the evolution equation as
\begin{equation}\label{eq:intpic}
	\partial_t\tilde{K}_P(t, \eta) = \left[\tilde{\mathcal{V}}(t)+ \tilde{\mathcal{J}}_{\eta}(t) + \tilde{\mathcal{W}}_{\eta}(t)\right]\tilde{K}_P(t, \eta).
\end{equation}
where we indicate the interaction picture with a tilde. To get the reduced dynamics we will introduce the Nakajima-Zwanzig projection operators defined by:
\begin{align}
	\mathcal{P}O & =\trR {O}\otimes \pi^{eq}_R,\\
	\mathcal{Q}O & =[\id-\mathcal{P}]O.
	\label{eq:proj2}
\end{align}
Due to the fact that $\trR{B_a \pi_R^{eq}}=0 \ \ \forall a$ , we have
\begin{align}
    &\mathcal{P} \tilde{\mathcal{W}}_\eta(t) = \tilde{\mathcal{W}}_\eta(t)\mathcal{P} \\
    &\mathcal{P} \tilde{\mathcal{J}}_{\eta}(t)\mathcal{P} = 0 \\
    &\mathcal{P} \tilde{\mathcal{V}}(t)\mathcal{P}=0
\end{align}
Now, applying the projection operators to~\eqref{eq:intpic}, and using these properties we have
\begin{align}
	\partial_t \mathcal{P}\tilde{K}_P(t, \eta) &= \mathcal{P}\left[\tilde{\mathcal{V}}(t)+ \tilde{\mathcal{J}}_{\eta}(t)\right]\mathcal{Q}\tilde{K}_P(t, \eta)+\tilde{\mathcal{W}}_{\eta}(t)\mathcal{P}\tilde{K}_P(t, \eta) \label{eq:wcoreduced},\\
	\partial_t \mathcal{Q}\tilde{K}_P(t, \eta) &= \left[\mathcal{Q}\left(\tilde{\mathcal{V}}(t)+ \tilde{\mathcal{J}}_{\eta}(t)\right) + \tilde{\mathcal{W}}_{\eta}(t) \right]\mathcal{Q}\tilde{K}_P(t, \eta)+\mathcal{Q}\left(\tilde{\mathcal{V}}(t)+\tilde{\mathcal{J}}_{\eta}(t)\right)\mathcal{P}\tilde{K}_P(t, \eta).
	\label{eq:proj2dif}
\end{align}
The solution to the homogeneous version of the second equation is 
\begin{equation}
	[\mathcal{Q}\tilde{K}_P(t, \eta)]_c = \tilde{G}(t,s) \mathcal{Q}\tilde{K}_P(0, \eta),
\end{equation}
where we have introduced the superpropagator
\begin{equation}
	\tilde{G}(t,s) =\overleftarrow{\mathcal{T}}\text{exp}\bigg(\int^t_s du\left[\mathcal{Q}\left(\tilde{\mathcal{V}}(u)+ \tilde{\mathcal{J}}_\eta(u)\right) + \tilde{\mathcal{W}}_\eta(u)\right]\bigg).
\end{equation}
To get the particular solution to the nonhomogeneous equation, we first multiply the complimentary solution from the right by an unknown operator $C(t)$:
\begin{equation}
	[\mathcal{Q}\tilde{K}_P(t, \eta)]_p = \tilde{G}(t,0) \mathcal{Q}\tilde{K}_P(0, \eta) C(t).
\end{equation}
Substituting this in to~\eqref{eq:proj2dif}, we get the differential equation  
\begin{equation}
\tilde{G}(t,0)\mathcal{Q}\tilde{K}_P(0, \eta)C'(t)=\mathcal{Q}\left(\tilde{\mathcal{V}}(t)+ \tilde{\mathcal{J}}_\eta(t)\right)\mathcal{P}\tilde{K}_P(t, \eta).
\end{equation}
We solve this equation for $C(t)$ to find the particular solution
\begin{equation}
	[\mathcal{Q}\tilde{K}_P(t, \eta)]_p = \int^t_0 ds \  \tilde{G}(t,s)\mathcal{Q}[\tilde{\mathcal{V}}(s)+ \tilde{\mathcal{J}}_\eta(s)]\mathcal{P}\tilde{K}_P(s, \eta),
\end{equation}
where we have used the composition rule for the superpropagator $\tilde{G}(t,s) = \tilde{G}(t,0)\tilde{G}^{-1}(s,0)$ and have chosen the arbitrary initial condition $C(0)=0$. The general solution to the nonhomogeneous equation is 
\begin{equation}
\begin{split}
	\mathcal{Q}\tilde{K}_P(t, \eta) &= [\mathcal{Q}\tilde{K}_P(t, \eta)]_c+[\mathcal{Q}\tilde{K}_P(t, \eta)]_p \\
	&= \tilde{G}(t,s) \mathcal{Q}\tilde{K}_P(0, \eta)+\int^t_0 ds \  \tilde{G}(t,s)\mathcal{Q}\left[\tilde{\mathcal{V}}(s)+ g\tilde{\mathcal{J}}_\eta(s)\right]\mathcal{P}\tilde{K}_P(s, \eta).
\end{split}
\end{equation}
We now choose the initial condition $\mathcal{Q}\tilde{K}_P(0, \eta) = 0$, which is true for an initial separable state, to get $\mathcal{Q}\tilde{K}_P(t, \eta) = \int^t_0 ds \  \tilde{G}(t,s)\mathcal{Q}[\tilde{\mathcal{V}}(s)+ \tilde{\mathcal{J}}_\eta(s)]\mathcal{P}\tilde{K}_P(s, \eta)$. Substituting this solution in to the first differential equation~\eqref{eq:wcoreduced} we have
\begin{equation}
	\partial_t \mathcal{P}\tilde{K}(t, \eta) = \tilde{\mathcal{W}}_\eta(t)\mathcal{P}\tilde{K}_P(t, \eta)+\mathcal{P}[\tilde{\mathcal{V}}(t)+ \tilde{\mathcal{J}}_\eta(t)]\int^t_0 ds \  \tilde{G}(t,s)\mathcal{Q}[\tilde{\mathcal{V}}(s)+ \tilde{\mathcal{J}}_\eta(s)]\mathcal{P}\tilde{\mathcal{U}}(s,t)\tilde{K}_P(t, \eta),
\end{equation}
where we have introduced 
\begin{equation}
	\tilde{\mathcal{U}}(s,t)\tilde{K}_P(t, \eta)= \tilde{U}(s,0) \ \overrightarrow{\mathcal{T}}\text{exp}\bigg[-\int^t_s du \  \tilde{U}^\dagger(u)\tilde{\mathcal{W}}_\eta(u)\tilde{U}(u)\bigg]\bigg( \tilde{U}^\dagger(t,0)\tilde{K}_P(t, \eta) \tilde{U}(t,0)\bigg)\tilde{U}^\dagger(s,0) ,
\end{equation}
to bring the equation into a time-convolutionless form, with global unitary 
\begin{align}\label{eq:unitary}
    U(t,t')=\overleftarrow{\mathcal{T}}\text{exp}\bigg(-i\int^{t'}_t ds \ \tilde{H}_P(s)\bigg)
\end{align}
in the interaction picture. If we rescale our interaction as $V_P\to g V_P$, such as we did in~equation 9 of the manuscript, then one can expand the propagators to zeroth order:
\begin{gather}
	\tilde{G}(t,s) = \overleftarrow{\mathcal{T}}\text{exp}\bigg(\int^t_s du \  \tilde{\mathcal{W}}_\eta(u)\bigg) + \mathcal{O}(g),\\
	\tilde{\mathcal{U}}(s,t) = \overrightarrow{\mathcal{T}}\text{exp}\bigg(-\int^t_s du \  \tilde{\mathcal{W}}_\eta(u)\bigg) + \mathcal{O}(g).
\end{gather}
Four our purposes this expansion is justified under the Markov approximation, which assumes that
\begin{equation}
	g \tau_B \ll 1,
	\label{eq:condition1}
\end{equation}
where $\tau_B$ is the characteristic time of the bath correlation function. Using these expansions we obtain a master equation for the WCO correct to second order in the system-bath coupling:
\begin{align}
	\nonumber \partial_t \mathcal{P}\tilde{K}_P(t, \eta) = \mathcal{P}[\tilde{\mathcal{V}}(t)+ \tilde{\mathcal{J}}_\eta(t)]\int^t_0 ds \  \overleftarrow{\mathcal{T}}\text{exp}\bigg(\int^t_s du \  \tilde{\mathcal{W}}_\eta(u)\bigg)[\tilde{\mathcal{V}}(s)+ \tilde{\mathcal{J}}_\eta(s)]\mathcal{P} \ \overrightarrow{\mathcal{T}}\text{exp}&\bigg(-\int^t_s du \  \tilde{\mathcal{W}}_\eta(u)\bigg)\tilde{K}_P(t, \eta) \\
	&+\tilde{\mathcal{W}}_\eta(t) \mathcal{P}\tilde{K}(t, \eta)+\mathcal{O}(g^2).
\end{align}
Explicitly writing the projection operators and $\tilde{\mathcal{V}}$ and $\tilde{\mathcal{J}}_\eta$, we obtain an equation for the \textit{reduced} WCO:
\begin{multline}\label{eq:2ndorder}
	\partial_t \tilde{K}_{PS}(t, \eta) -\tilde{\mathcal{W}}_\eta(t)\tilde{K}_{PS}(t, \eta)
	=\sum_{a,b} \int^t_0 ds\big[ e^{i\eta \tilde{H}_{PS}(t)}\tilde{A}^\dagger_a(t)e^{-i\eta \tilde{H}_{PS}(t)} \tilde{K}_{PS}(t, \eta) \tilde{A}_b(t-s)\big\<\tilde{B}_b(-s-\eta)\tilde{B}_a(0)\big>\\
	-\bigg(e^{i \eta \tilde{H}_{PS}(t)}\tilde{A}^\dagger_a(t)e^{-i \eta \tilde{H}_{PS}(t)}\overleftarrow{\mathcal{T}}\text{exp}\bigg(\int^t_{t-s} du \tilde{\mathcal{W}}_\eta(u)\bigg) e^{i \eta \tilde{H}_{PS}(t-s)}\tilde{A}_b(t-s)e^{-i \eta \tilde{H}_{PS}(t-s)} \\
 \times\overrightarrow{\mathcal{T}}\text{exp}\bigg(-\int^t_{t-s} du \tilde{\mathcal{W}}_\eta(u)\bigg)\tilde{K}_{PS}(t, \eta)\big\<\tilde{B}_a(s)\tilde{B}_b(0)\big\>\bigg)
	-\tilde{K}_{PS}(t, \eta) \tilde{A}_b(t-s) \tilde{A}^\dagger_a(t)\big\<\tilde{B}_b(-s)\tilde{B}_a(0)\big\>\\
	+\overleftarrow{\mathcal{T}}\text{exp}\bigg(\int^t_{t-s} du \tilde{\mathcal{W}_\eta}(u)\bigg) e^{i \eta \tilde{H}_{PS}(t-s)}\tilde{A}_b(t-s)e^{-i \eta \tilde{H}_{PS}(t-s)}\overrightarrow{\mathcal{T}}\text{exp}\bigg(-\int^t_{t-s} du \tilde{\mathcal{W}}_\eta(u)\bigg)\tilde{K}_{PS}(t, \eta)A_a^\dagger(t) \big\<\tilde{B}_a(s-\eta)\tilde{B}_b(0)\big\>\Big].
\end{multline}
At this stage we will make an adiabatic approximation, following the approach of \citeauthor{albash2012} \cite{albash2012}, and assume that the system Hamiltonian is varied sufficiently slowly. If we decompose the polaron system Hamiltonian in terms of its energy eigenstates, $H_{PS}(t)=\sum_n \epsilon_n(t)\ket{\epsilon_n(t)}\bra{\epsilon_n(t)}$, then the standard condition for adiabaticity is 
\begin{equation}
	\mathcal{A} = \max_{t\in[0, t_f], n \neq m} \left|\frac{\bra{\epsilon_n(t)}{\partial_t H_{PS}(t)}\ket{\epsilon_m(t)}}{\omega_{nm}^2(t)}\right| \ll 1.
	\label{eq:adiabatic.condition}
\end{equation}
which means that the rate of change of the eigenbasis is small relative to the energy gaps of the polaron system Hamiltonian. We also require the rate of change of the eigenbasis,
\begin{equation}
	h = \max_{t\in[0, t_f], n \ne m} |\bra{\epsilon_n(t)}{\partial_t H_{PS}(t)}\ket{\epsilon_m(t)}|,
\end{equation}
to be small relative to the bath characteristic timescale $\tau_B$, so we have
\begin{equation}
	h \tau_B^2 \ll 1
	\label{eq:condition2}.
\end{equation}
If these conditions hold true then the system interaction operators in the interaction picture become
\begin{align}		
	\tilde{A}_a(t) &=\sum_{n,m} A_{a,nm}^\dagger(t) e^{i \mu_{nm}(t)}=\sum_{n,m} A_{a,mn}(t)  e^{-i \mu_{mn}(t)},\\
	\tilde{A}_a(t-s) &=\sum_{n,m} A_{a,nm}^\dagger(t) e^{i \mu_{nm}(t)-is\omega_{nm}(t)}=\sum_{n,m} A_{a,mn}(t)  e^{-i \mu_{mn}(t)+is\omega_{mn}(t)},
\end{align}
where
\begin{align}
	A_{a,nm}^\dagger(t) & = \bra{\epsilon_n(t)}A_a\ket{\epsilon_m(t)}\ket{\epsilon_n(0)}\bra{\epsilon_m(0)}=A_{a,mn}(t),\\
	\mu_{nm}(t) &= \mu_n(t)-\mu_m(t),\\
	\omega_{mn}(t)&=\epsilon_m(t)-\epsilon_n(t).
\end{align}
and 
\begin{equation}
	\mu_n(t) = \int^t_0 ds [\epsilon_n(s)-i \<\epsilon_n(s)|\dot{\epsilon}_n(s)\>],
	\label{eq:adphase}
\end{equation}
is the time-integral of the sum of the adiabatic and geometric phases. From these definitions one can derive the following commutators:
\begin{align}
	[\tilde{H}_{PS}(t),\tilde{A}_{a,mn}(t)] &= -\omega_{mn}(t)\tilde{A}_{a,mn}(t),\label{eq:eigenop3}\\
	[\tilde{H}_{PS}(t),\tilde{A}^\dagger_{a,mn}(t)] &= \omega_{mn}(t)\tilde{A}^\dagger_{a,mn}(t).\label{eq:eigenop4}
\end{align} 
This implies 
\begin{align}
	\tilde{A}_{a,mn}(t)e^{-i\eta \tilde{H}_{PS}(t)}&=e^{-i \eta \omega_{nm}(t)}e^{-i\eta \tilde{H}_{PS}(t)}\tilde{A}_{a,mn}(t),	\label{eq:eigenop.ad1}\\
	\tilde{A}^\dagger_{a,mn}(t)e^{-i\eta \tilde{H}_{PS}(t)}&=e^{i \eta \omega_{nm}(t)}e^{-i\eta \tilde{H}_{PS}(t)}\tilde{A}^\dagger_{a,mn}(t),
	\label{eq:eigenop.ad2}
\end{align}
We can use these identities along with our approximations of the interaction operators to evaluate the RHS of~\eqref{eq:2ndorder}:
RHS may be written as 
\begin{multline*}
	\sum_{a,b,n,m,k,l} \int^t_0 e^{i[\mu_{nm}(t)-\mu_{lk}(t)+s \omega_{lk}(t)]} e^{i\eta \tilde{H}_{PS}(t)} ds\Big[ \tilde{A}^\dagger_{a,nm}(t)e^{-i\eta \tilde{H}_{PS}(t)} \tilde{K}_{PS}(t,\eta) \tilde{A}_{b,lk}(t)\mathcal{B}_{ba}(-s-\eta)\\
	-\tilde{A}^\dagger_{a,nm}(t)\tilde{A}_{b,lk}(t)e^{-i\eta \tilde{H}_{PS}(t)}\tilde{K}_{PS}(t,\eta)\mathcal{B}_{ab}(s)
	-\tilde{K}_{PS}(t, \eta) \tilde{A}_{b,lk}(t) \tilde{A}^\dagger_{a,nm}(t)\mathcal{B}_{ab}(s)\\
	+\tilde{A}_{b,lk}(t)e^{-i\eta \tilde{H}_{PS}(t)}\tilde{K}_{PS}(t,\eta)\tilde{A}^\dagger_{a,nm}(t)\mathcal{B}_{ab}(s-\eta)\Big],
\end{multline*}
where we introduce the bath correlation function 
\begin{align}
    \mathcal{B}_{ab}(t)=\big\<\tilde{B}_a(t)\tilde{B}_b(0)\big\>,
\end{align}
We can pass the term $e^{-i \eta \tilde{H}_{PS}(t)}$ past the system interaction operators using~\eqref{eq:eigenop.ad2} to get
\begin{multline*}
	\sum_{a,b,n,m,k,l} \int^t_0 ds \ e^{i[\mu_{nm}(t)-\mu_{lk}(t)+s \omega_{lk}(t)]}  \Big[ \tilde{A}^\dagger_{a,nm}(t) \tilde{K}_{PS}(t,\eta) \tilde{A}_{b,lk}(t)e^{i\eta \omega_{nm}(t)}\mathcal{B}_{ba}(-s-\eta)\\
	-\tilde{A}^\dagger_{a,nm}(t)\tilde{A}_{b,lk}(t)\tilde{K}_{PS}(t,\eta)e^{i\eta [\omega_{nm}(t)-\omega_{lk}(t)]}\mathcal{B}_{ab}(s)
	-\tilde{K}_{PS}(t, \eta) \tilde{A}_{b,lk}(t) \tilde{A}^\dagger_{a,nm}(t)\mathcal{B}_{ba}(-s)\\
	+\tilde{A}_{b,lk}(t)\tilde{K}_{PS}(t,\eta)\tilde{A}^\dagger_{a,nm}(t)e^{-i\eta\omega_{lk}(t)}\mathcal{B}_{ab}(s-\eta)\Big].
\end{multline*}
Again following \citeauthor{albash2012}, we make a \textit{secular-like approximation}, arguing that in the $t\to\infty$ limit the dominant terms will be those where $\mu_{nm}(t)-\mu_{lk}(t)=0$, which from~\eqref{eq:adphase} occurs when $n=m$ and $l=k$ or when $n=l$ and $m=k$. Here the integrated adiabatic phases $\mu_{nm}$ mirror the static Bohr frequencies that we make the standard secular approximation on. Removing the non-secular terms we see that the RHS is equal to the sum of two terms. The first term is 
\begin{multline}
	\sum_{a,b,n,m} \Big[ \tilde{A}^\dagger_{a,nn}(t) \tilde{K}_{PS}(t,\eta) \tilde{A}_{b,mm}(t)\int^\infty_0 ds \ \mathcal{B}_{ab}(s+\eta)
	-\tilde{A}^\dagger_{a,nn}(t)\tilde{A}_{b,mm}(t)\tilde{K}_{PS}(t,\eta)\int^\infty_0 ds \ \mathcal{B}_{ab}(s)\\
	-\tilde{K}_{PS}(t, \eta) \tilde{A}_{b,mm}(t) \tilde{A}^\dagger_{a,nn}(t)\int^\infty_0 ds \ \mathcal{B}_{ba}(-s)
	+\tilde{A}_{b,mm}(t)\tilde{K}_{PS}(t,\eta)\tilde{A}^\dagger_{a,nn}(t)\int^\infty_0 ds \ \mathcal{B}_{ab}(s-\eta)\Big],
\end{multline}
and the second term is
\begin{multline}
	\sum_{a,b,n\ne m} \Big[ \tilde{A}^\dagger_{a,nm}(t) \tilde{K}_{PS}(t,\eta) \tilde{A}_{b,nm}(t)\int^\infty_0 ds \ e^{i(s-\eta)\omega_{nm}(t)}\mathcal{B}_{ab}(s-\eta)-\tilde{A}^\dagger_{a,nm}(t)\tilde{A}_{b,nm}(t)\tilde{K}_{PS}(t,\eta)\int^\infty_0 ds \ e^{is\omega_{nm}(t)}\mathcal{B}_{ab}(s)\\
	-\tilde{K}_{PS}(t, \eta) \tilde{A}_{b,nm}(t) \tilde{A}^\dagger_{a,nm}(t)\int^\infty_0 ds \ e^{is\omega_{nm}(t)}\mathcal{B}_{ba}(-s)+\tilde{A}_{b,nm}(t)\tilde{K}_{PS}(t,\eta)\tilde{A}^\dagger_{a,nm}(t)\int^\infty_0 ds \ e^{i(s-\eta)\omega_{nm}(t)}\mathcal{B}_{ab}(s-\eta)\Big].
	\label{eq:sum}
\end{multline}
The integrals over the bath correlation function may be rewritten as Fourier transforms (denoted by operation $\mathcal{F}(.)$). For example, the last integral of~\eqref{eq:sum} can be expressed as 
\begin{equation}
	e^{-i\eta\omega}\int^\infty_0 ds \ e^{is\omega}\mathcal{B}_{ab}(s-\eta) =e^{-i\eta\omega}\mathcal{F}[\mathcal{B}_{ab}(s-\eta)\Theta(s)]= e^{-i\eta\omega}\mathcal{F}[\mathcal{B}_{ab}(s-\eta)]\ast\mathcal{F}[\Theta(s)].
\end{equation}
where $\Theta(s)$ denotes the step function. The first Fourier transform can be expressed in terms of the Hermitian part of the spectral density matrix 
\begin{equation}
	\mathcal{F}[\mathcal{B}_{ab}(s-\eta)] = e^{i \omega \eta}\gamma_{ab}(\omega), \ \ \ \ \ \ \ \gamma_{ab}(\omega) = \int^\infty_{-\infty}d \tau \  e^{i \omega \tau}\mathcal{B}_{ab}(\tau),
\end{equation}
and for the second we can use $\mathcal{F}[\Theta(s)] = \text{p.v.} \frac{1}{i\omega} + \pi \delta(\omega)$. Using these results we have 
\begin{equation}
	e^{-i\eta\omega}\int^\infty_0 ds \ e^{is\omega}\mathcal{B}_{ab}(s-\eta) = \frac12 \gamma_{ab}(\omega) -i S^\eta_{ab}(\omega),
\end{equation}
where we have defined the counting-field shifted non-Hermitian part of the spectral density matrix:
\begin{align}
	S^\eta_{ab}(\omega)  = \frac1{2\pi}\text{p.v.}\int^\infty_{-\infty} d\omega' \frac{\gamma_{ab}(\omega')}{\omega-\omega'}e^{-i (\omega -\omega')\eta}=S_{ba}^{-\eta}(\omega)^\ast.
	\label{eq:parts}
\end{align}
In the second equality we used the property of the correlation function $\mathcal{B}_{ab}(\tau)=\mathcal{B}_{ba}(-\tau)^\ast$. Using the same procedure we find 
\begin{gather}
	e^{i\eta\omega}\int^\infty_0 ds \ e^{is\omega}\mathcal{B}_{ba}(-s-\eta) = \frac12 \gamma_{ab}(-\omega)^\ast -i S^{-\eta}_{ab}(-\omega)^\ast,\\
	\int^\infty_0 ds \ e^{is\omega}\mathcal{B}_{ab}(s) = \frac12 \gamma_{ab}(\omega) -i S_{ab}(\omega),\\
	\int^\infty_0 ds \ e^{is\omega}\mathcal{B}_{ba}(-s) = \frac12 \gamma_{ab}(-\omega)^\ast -i S_{ab}(-\omega)^\ast.
\end{gather}
where we denoted $S_{ab}(\omega)=S^{\eta=0}_{ab}(\omega)$. Considering the final term in the sum in equation~\eqref{eq:sum}, and inserting the expressions for the integrals,
\begin{align}
\nonumber&\sum_{\substack{a,b\\n\ne m}} \tilde{A}^\dagger_{a,nm}(t) \tilde{K}_{PS}(t,\eta) \tilde{A}_{b,nm}(t)\bigg(\frac{\gamma_{ab}[\omega_{nm}(t)]}2  -i S^\eta_{ab}[\omega_{nm}(t)]\bigg) \\
\nonumber & \ \ \ \ \ \ \ \ \ \ \ \ \ \ \ \ \ \ \ \ \ \ \ \ \ \ \ \ \ \ \ \ \  +\tilde{A}_{b,nm}(t)\tilde{K}_{PS}(t,\eta)\tilde{A}^\dagger_{a,nm}(t)\bigg(\frac{\gamma_{ab}[-\omega_{nm}(t)]^\ast}2  -i S^{-\eta}_{ab}[-\omega_{nm}(t)]^\ast\bigg)\\
	\nonumber&=\sum_{\substack{a,b\\n\ne m}} \tilde{A}^\dagger_{a,nm}(t) \tilde{K}_S(t,\eta) \tilde{A}_{b,nm}(t)\bigg(\frac{\gamma_{ab}[\omega_{nm}(t)]}2  -i S^\eta_{ab}[\omega_{nm}(t)]\bigg) \\
\nonumber & \ \ \ \ \ \ \ \ \ \ \ \ \ \ \ \ \ \ \ \ \ \ \ \ \ \ \ \ \ \ \ \ \ +\tilde{A}_{a,mn}(t)\tilde{K}_{PS}(t,\eta)\tilde{A}^\dagger_{b,mn}(t)\bigg(\frac{\gamma_{ba}[-\omega_{mn}(t)]^\ast}2  -i S^{-\eta}_{ba}[-\omega_{mn}(t)]^\ast\bigg)\\
	\nonumber&=\sum_{\substack{a,b\\n\ne m}} \tilde{A}^\dagger_{a,nm}(t) \tilde{K}_{PS}(t,\eta) \tilde{A}_{b,nm}(t)\bigg(\frac{\gamma_{ab}[\omega_{nm}(t)]}2  -i S^\eta_{ab}[\omega_{nm}(t)]\bigg) \\
\nonumber & \ \ \ \ \ \ \ \ \ \ \ \ \ \ \ \ \ \ \ \ \ \ \ \ \ \ \ \ \ \ \ \ +\tilde{A}^\dagger_{a,nm}(t) \tilde{K}_{PS}(t,\eta) \tilde{A}_{b,nm}(t)\bigg(\frac{\gamma_{ba}[\omega_{nm}(t)]^\ast}2  -i S^{-\eta}_{ba}[\omega_{nm}(t)]^\ast\bigg) \\
	&=\sum_{\substack{a,b\\n\ne m}} \gamma_{ab}[\omega_{nm}(t)] \tilde{A}^\dagger_{a,nm}(t) \tilde{K}_{PS}(t,\eta) \tilde{A}_{b,nm}(t),
\end{align}
where in the first equality we swapped indices $a\leftrightarrow b$ and $n\leftrightarrow m$, in the second we used the identities $\tilde{A}_{a,nm}=\tilde{A}_{a,mn}^\dagger$ and $\omega_{nm}=-\omega_{mn}$, and in the last we used the identities in~\eqref{eq:parts} to cancel the principal value integrals. Repeating this procedure with the other terms we find that
\begin{equation}
	\partial_t\tilde{K}_{PS}(t,\eta) = \tilde{\mathcal{L}}_t(\tilde{K}_{PS}(t,\eta))+[\partial_t e^{i \eta \tilde{H}_{PS}(t)}]e^{-i \eta \tilde{H}_{PS}(t)}\tilde{K}_{PS}(t,\eta),
\end{equation}
or in the Schr{\"o}dinger picture 
\begin{equation}
	\partial_t K_{PS}(t,\eta) = \mathcal{L}_t[K_{PS}(t,\eta)]+\mathcal{W}_\eta(t) K_{PS}(t,\eta),
\end{equation}
where $\mathcal{L}_t$ is the adiabatic Lindbladian:
\begin{multline}	
	\mathcal{L}_t[.]=-i [H_{PS}(t)+H_{PLS}(t),(.)]+\sum_{a,b}\sum_{n \ne m} \gamma_{a b}(\omega_{mn}(t))\left[A_{nm, b}(t)(.) A_{nm, a}^\dagger(t) -\frac12 \big\{A_{nm, a}^\dagger(t) A_{nm, b}(t),(.)\big\}\right]\\
	+\sum_{a, b}\sum_{n,m} \gamma_{a b}(0)\left[A_{nn, b}(t)(.) A_{mm, a}^\dagger(t) -\frac12 \big\{A_{nn, a}^\dagger(t) A_{mm, b}(t),(.)\big\}\right]
	\label{eq:ame},
\end{multline}
where the Lamb shift Hamiltonian is 
\begin{equation}
	H_{LS}(t) = \sum_{a, b}\sum_{n\ne m} A_{nm, a}^\dagger(t)A_{nm, b}(t) S_{ab}(\omega_{mn}(t))+\sum_{a, b}\sum_{n,m} A^\dagger_{nn, a}(t)A_{mm, b}(t)S_{a b}(0).
\end{equation}
The jump operators are 
\begin{equation}
	A_{nm, a}(t) = A_{mn, a}^\dagger(t) = \bra{\epsilon_n(t)}A_a\ket{\epsilon_m(t)}\ket{\epsilon_n(t)}\bra{\epsilon_m(t)},
\end{equation}
This concludes the derivation of the WCO master equation presented in the main text. In the next section we evaluate the specific jump operators and correlation functions for a polaron system.

\section{Polaron versus weak-coupling adiabatic master equations }\label{appB}

\

\noindent In this appendix we provide detailed expressions for the dissipator and renormalisation factor used to model the dynamics of the system and its resulting work statistics. For comparison, we present the master equations obtained through (1) the polaron framework and (2) the weak coupling approximation.

\subsection{The polaron adiabatic master equation}

As stated in the main text, the interaction Hamiltonian in the polaron frame is given by
\begin{align}
    \nonumber&V_P=\frac{\Delta}{2} \big(\sigma_x\otimes\xi_x+\sigma_y\otimes\xi_y\big), \\
    \nonumber&\xi_x=\frac{1}{2}\big(\xi_+ +\xi_{-}-\kappa\big), \\
    &\xi_y=\frac{i}{2}\big(\xi_+ -\xi_{-}\big),
\end{align}
where $\kappa$ is the polaron renormalisation factor and $\xi_{\pm}=\prod_k D(\pm \alpha_k)$ is a product of displacement operators, where the $k$th mode is displaced by $\alpha_k=g_k/\omega_k$ with $\omega_k$ the frequency of the mode and $g_k$ its corresponding coupling constant. The bath is characterised by a cubic spectral density of the form
\begin{align}
    J(\omega)= \alpha\bigg(\frac{ \omega^3}{\omega_c^2}\bigg)e^{-\omega/\omega_c},
\end{align}
To obtain the bath correlation function, one may show
\begin{align}
	\big\<\xi_\pm(t)\xi_\pm(0)\big\> & = \kappa^2 e^{-\phi(t)},\\
	\big\<\xi_\pm(t)\xi_\mp(0)\big\> & = \kappa^2 e^{\phi(t)},
\end{align}
where
\begin{align}
	\nonumber \phi(t) &= 4\int^\infty_0 d\omega  \frac{J(\omega)}{\omega^2} \left(\coth(\beta \omega/2) \cos \omega t -i \sin \omega t \right) , \\
	&= -4\alpha\left[(1-i\omega_c t)^{-2}-\epsilon^2[\psi^{(1)}(\epsilon+it/\beta)+\psi^{(1)}(\epsilon-i t/\beta)]\right]
\end{align}
is the bath propagator, where we have defined $\epsilon=1/\beta \omega_c$ and where $\psi^{(1)}(x)$ is the first polygamma function. The polaron normalisation constant evaluates to
\begin{align}
    \nonumber\kappa&=\exp\bigg(-2 \int^\infty_0 d\omega \frac{J(\omega)}{\omega^2}\coth \big(\beta\omega/2\big)\bigg), \\
    &=\text{exp}\bigg(-2\alpha \big(2\epsilon^2 \psi^{(1)}(\epsilon)-1\big)\bigg)
\end{align}
The correlation functions for the interaction terms are
\begin{align}
	\big\<\xi_x(t)\xi_x(0)\big\> & =  [\cosh \phi(t) -1],\\
	\big\<\xi_y(t)\xi_y(0)\big\> & = \kappa^2 \sinh \phi(t),\\
	\big\<\xi_x(t)\xi_y(0)\big\> & = \big\<\xi_y(t)\xi_x(0)\big\>=0.
\end{align}
With these we can evaluate the hermitian part of the spectral density function:
\begin{align}
    &\gamma_{xx}(\omega) = \kappa^2\int^\infty_{-\infty}d \tau \  e^{i \omega \tau}[\cosh \phi(t) -1], \\
    &\gamma_{yy}(\omega) = \kappa^2\int^\infty_{-\infty}d \tau \  e^{i \omega \tau}\sinh \phi(t),
\end{align}
and the imaginary part
\begin{align}
    &S_{xx}(\omega)=\frac1{2\pi}\text{p.v.}\int^\infty_{-\infty} d\omega' \frac{\gamma_{xx}(\omega')}{\omega-\omega'} \\
    &S_{yy}(\omega)=\frac1{2\pi}\text{p.v.}\int^\infty_{-\infty} d\omega' \frac{\gamma_{yy}(\omega')}{\omega-\omega'}
\end{align}
\noindent These integrals can be handled with standard numerical integration methods. The eigenstates of the polaron frame system Hamiltonian are
\begin{align}
	\ket{\varepsilon_+(t)}&=\cos\frac{\theta(t)}2\ket{1} + \sin\frac{\theta(t)}2 \ket{0},\\
	\ket{\varepsilon_-(t)}&=-\sin\frac{\theta(t)}2 \ket{1}+ \cos\frac{\theta(t)}2 \ket{0},
\end{align}
where $\theta(t) = \arctan(\kappa \Delta/\omega_0(t))$, and the energy eigenvalues are $\varepsilon_\pm(t) = \pm\frac12\omega(t)$, with $\omega(t) = \sqrt{\omega_0^2(t)+\kappa^2 \Delta^2}$ the transition frequency. We now write the adiabatic Lindblad generator~\eqref{eq:ame} in the form
\begin{align}
    \mathcal{L}_{t}(.)=-i[H_{PS}(t)+H_{PLS}(t),(.)]+ \mathcal{D}_t(.)
\end{align}
where the dissipator is expressed more compactly as:
\begin{equation}\label{eq:PMEfinal}
	\mathcal{D}_t(.) = \sum_{n=\pm,0} \ \sum_{a=x, y} \gamma_{a, n}(t) \left(A_{a,n}(t)(.) A_{a,n}^\dagger(t) - \frac12 \{A_{a,n}(t)A^\dagger_{a,n}(t),(.)\}\right),
\end{equation}
Here we have defined $\gamma_{a, \pm}(t) \equiv \gamma_{aa}(\mp\omega(t))$ and  $\gamma_{a, 0} \equiv \gamma_{aa}(0)$. The jump operators are now
\begin{align}
	A_{a, +}(t) = A_{a, -}^\dagger(t) &= \bra{\varepsilon_+(t)}A_a\ket{\varepsilon_-(t)}\ket{\varepsilon_+(t)}\bra{\varepsilon_-(t)},\\
	A_{a, 0}(t) & = \bra{\varepsilon_+(t)}A_a\ket{\varepsilon_+(t)}\ket{\varepsilon_+(t)}\bra{\varepsilon_+(t)}-\bra{\varepsilon_-(t)}A_a\ket{\varepsilon_-(t)}\ket{\varepsilon_-(t)}\bra{\varepsilon_-(t)},
\end{align}
and then using the definitions of the eigenstates and interaction Hamiltonian, 
\begin{align}
	A_{x, +}(t) = A_{x, -}^\dagger(t) &= \frac{\Delta}{2}\cos\theta(t)\ket{\varepsilon_+(t)}\bra{\varepsilon_-(t)},\\
	A_{a, 0}(t) & = \frac{\Delta}{2}\sin\theta(t)\left(\ket{\varepsilon_+(t)}\bra{\varepsilon_+(t)}-\ket{\varepsilon_-(t)}\bra{\varepsilon_-(t)} \right),\\
	A_{y, +}(t) = A_{y, -}^\dagger(t) &= - \frac{i\Delta}{2}\ket{\varepsilon_+(t)}\bra{\varepsilon_-(t)},\\
	A_{y, 0}(t) & = 0.
\end{align}
The Lamb shift Hamiltonian is 
\begin{equation}\label{eq:lamb}
	H_{PLS}(t) =  \sum_{n=\pm, 0} \ \sum_{a=x, y} S_{a,n}(t) A^\dagger_{a,n}(t) A_{a,n}(t),
\end{equation}
where we have defined $S_{a,\pm}(t)\equiv S_{aa}(\mp\omega(t))$ and $S_{a,0}\equiv S_{aa}(0)$. 

\subsection{Weak coupling master equation}

\

From the previous section we have presented the polaron master equation (PME)~\eqref{eq:PMEfinal} and found that it drives transitions between instantaneous eigenstates of the polaron system Hamiltonian $H_{PS}(t)$. In the main text we compare this with simulations of the standard adiabatic weak-coupling master equation (WCME), which instead drives transitions between eigenstates of the original Hamiltonian $H_S(t)$. These eigenstates are given by
\begin{align}
	\ket{\varepsilon_+(t)}&=\cos\frac{\theta(t)}2\ket{1} + \sin\frac{\theta(t)}2 \ket{0},\\
	\ket{\varepsilon_-(t)}&=-\sin\frac{\theta(t)}2 \ket{1}+ \cos\frac{\theta(t)}2 \ket{0},
\end{align}
 with $\theta(t) = \arctan(\Delta/\nu t)$ and energy eigenvalues  $\varepsilon_\pm(t) = \pm\frac12\omega(t)$ with $\omega(t) = \sqrt{\nu^2 t^2+ \Delta^2}$ the transition frequency. A time-dependent master equation can be derived under the weak coupling and adiabatic approximations using the approach of \cite{albash2012}. In contrast with the PME, there are only eigenoperators appearing in the WCME are $A_+(t) = A_-^\dagger(t) = \sin \theta(t)\dyad{\varepsilon_+(t)}{\varepsilon_-(t)}$ due to the fact that the interaction couples only through $\sigma_x$ in the original frame. The respective Lindblad rates are
\begin{align}
	\gamma_-(t) &=2\pi J(\omega(t))(1+N(\omega(t))),\\
	\gamma_+(t) &= 2\pi J(\omega(t))N(\omega(t)),
\end{align}
where the dephasing rate $\gamma_0(t)$ is zero because we consider a super-Ohmic spectral density, and $N(\omega)=1/(e^{\beta\omega}-1)$ the Bose-Einstein distribution. The WCME is then
\begin{gather}\label{eq:WCME}
		\dv{t}\rho_S(t) = -i \comm*{H_S(t)+H_{LS}(t)}{\rho_S(t)} + \mathcal{D}_t(\rho(t)) ,\\
	\mathcal{D}_t(\rho_S(t)) = \sum_{n=+,-} \gamma_n(t) [ A_n(t) \rho_S(t) A_n^\dagger(t) - \frac12 \acomm*{A_n^\dagger(t) A_n(t)}{\rho_S(t)}] \\
	H_{LS}(t) = \sum_{n=+,-} S_{n}(t) A^\dagger_{n}(t) A_{n}(t),
\end{gather}
and $S_{\pm}(t)=S_{x,\pm}(t)$ as defined in~\eqref{eq:lamb}.

\section{Parameter regimes for the Landau-Zener model}\label{app:parameter-regimes}

Here we discuss the valid parameter ranges in which we can use our master equation to obtain work statistics in the Landau-Zener model. First we consider the Born-Markov approximation in the polaron frame, which amounts to assuming that the environment can respond instantaneously to the dynamics of the system, where the dynamical timescale of the reservoir is proportional to the cutoff frequency $\omega_c$. 
From the main text recall the definition of the interaction strength by $g:=\frac{\Delta}{2}\sqrt{(1+\kappa^4)/2}$, which requires $g/\omega_c\ll1$ \cite{mccutcheon2011}. So long as we have a sufficiently large cutoff frequency $\Delta\ll \omega_c$, this condition can be satisfied from low to high temperatures, and from weak to strong coupling as defined by the original frame (small to large $\alpha$). This provides us with a much wider parameter regime to probe the work statistics than would be possible with the WCME.  Second, to ensure that the adiabatic approximation holds for the time-dependent protocol, we need the rate of change of the polaron's
eigenbasis to be small relative to the energy gaps of its Hamiltonian. This is achieved when 
\begin{equation}\label{eq:adiabatic_cond}
	\max_{t \in [t_i, t_f]}\frac{\Delta \kappa \nu}{2\omega(t)^{3}}=\frac{\nu}{2\Delta^2 \kappa^2}\ll 1,
\end{equation}
where $t_i$ and $t_f$ are the initial and final times respectively \cite{yamaguchi2017markovian}. The maximum is found at the intermediate avoided crossing where $t=0$. If we compare this condition with the  Born-Markov condition, we see that this can be maintained at any temperature so long as we choose a large enough cutoff frequency $\omega_c$. 
As a final condition we also require that the rate of change of the polaron energy eigenbasis must be small relative to the reservoir dynamics \cite{albash2012,mozgunov2020,dai2022,Chen2022}, which means $\nu\ll 2\omega_c$; this is guaranteed if the previous conditions are satisfied.  To confirm that polaron theory gives a faithful description of the reduced system, we have benchmarked the system population dynamics using the numerically exact TEMPO method~\cite{strathearn2018efficient} in the Supplementary Material and find excellent agreement with the adiabatic polaron dynamics at strong coupling, in contrast to the WCME. Therefore, we conclude that in the Landau-Zener model our master equation for the WCO, Eq.~(9) of the manuscript, is accurate so long as $\Delta\ll \omega_c$ and $\nu\ll 2\Delta^2\kappa^2$. We should emphasise that the parameters we have considered fall within experimentally relevant regimes. Taking the parameter $\Delta$ to be of the typical order of $10$~$\mu$eV, our presented results span a temperature range of order $100$~mK to $1$~K, reorganisation energies ranging from sub $\mu$eV to a few tens of $\mu$eV, and a cutoff $\omega_c\sim100$~$\mu$eV, all of which are reasonable experimental values for solid-state systems such as a double quantum dot.

\begin{figure}[t!]
    \centering
    \includegraphics[width=0.6\textwidth]{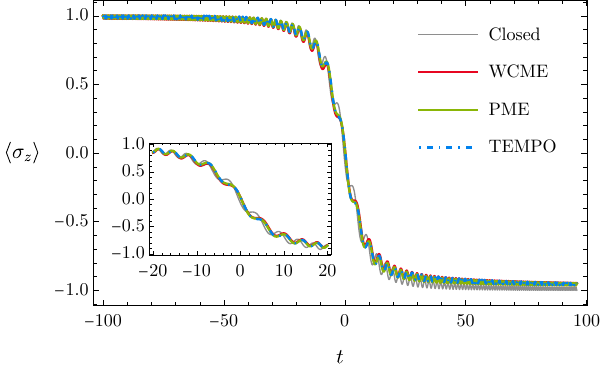}
    \includegraphics[width=0.6\textwidth]{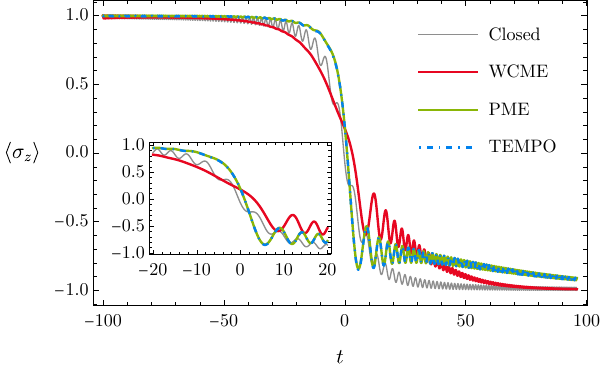}
    \includegraphics[width=0.6\textwidth]{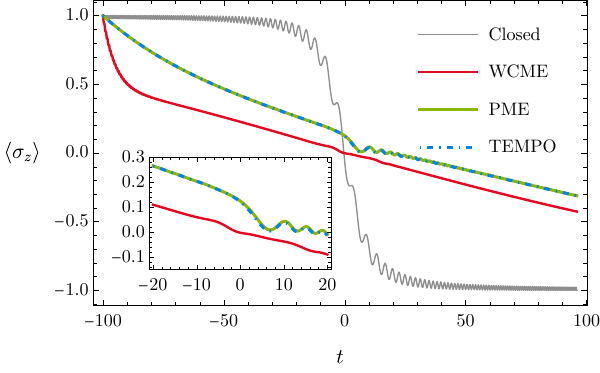}
    \caption{Population dynamics of the dissipative Landau-Zener model predicted by different dynamical models; closed unitary evolution, the weak-coupling master equation (WCME)~\eqref{eq:WCME}, the polaron master equation (PME)~\eqref{eq:PMEfinal} and an exact numerical simulation provided by TEMPO. The insets shows the dynamics near to the avoided crossing. The top and middle plots  correspond to a system-environment coupling strength of $\alpha=0.02$ and $\alpha=0.4$ respectively, both with $\Delta\beta=1$. 
    The bottom plot shows the dynamics for $\alpha=0.4$ and $\Delta\beta=0.1$.  Other parameters used: $\nu=0.1\Delta^2$, $t_0=-100/\Delta$, $t_f=100/\Delta$, and  $\omega_c=10\Delta$. The time axis is in units of $1/\Delta$. \label{fig:tempo}}
\end{figure}

\section{Benchmarking}\label{appC}

Details of a simulation of the time-dependent LZ model to justify our choice of parameters used to compute the work statistics in Fig.~1 and Fig.~2 of the main manuscript are provided here. We benchmark the reduced dynamics predicted by the PME~\eqref{eq:PMEfinal} with numerically exact dynamics calculated with the TEMPO algorithm \cite{strathearn2018efficient,gribben2022exact}. In Fig.~\ref{fig:tempo} below we show the population dynamics in the diabatic basis $\expval{\sigma_z} = \tr_S(\sigma_z \rho_S(t))=\tr_S(\sigma_z \rho_{PS}(t))$ with an initial state $\rho_0 = \dyad{1}$. We see excellent agreement between PME and TEMPO calculated dynamics, whilst the WCME~\eqref{eq:WCME} significantly overestimates the effect of the bath.  
For the tempo simulations, convergence was achieved with an singular value decomposition cutoff of $10^{-7}$, timestep of $\delta t= 0.025$, and memory cutoff of $\tau_c=2$.
Before starting the protocol, we allowed the bath to  relax to a displaced thermal state, using a time window of $\tau_W=4$. This means that the protocol is initialised with the bath and system in a true equilibrium state of the system and environment.

\section{Average and fluctuations of the work done}\label{app:work-moments}
We can use the work distributions to investigate
how the increased likelihood of stochastic dissipation of the system induced by strong reservoir coupling impacts both the average and fluctuations in the work done. 
A comparison of the the average work $\expval{W}$ and work variance $\operatorname{var}(W)=\expval{W^2}-\expval{W}^2$ calculated by the generalised PME and WCME as functions of the coupling for various inverse temperatures is shown in Fig~\ref{sup:grid} below. 
Since there is no change in the free-energy of the system-bath from the beginning to end of the protocol, any work done is dissipative. 
Here we see that the weak coupling theory typically underestimates both the average work and its variance for the range of temperatures and coupling strengths considered. The polaron and weak coupling predictions only coincide in the limit of very weak coupling. 

\begin{figure}
  \centering
  \subfloat{\includegraphics[width=0.425\textwidth]{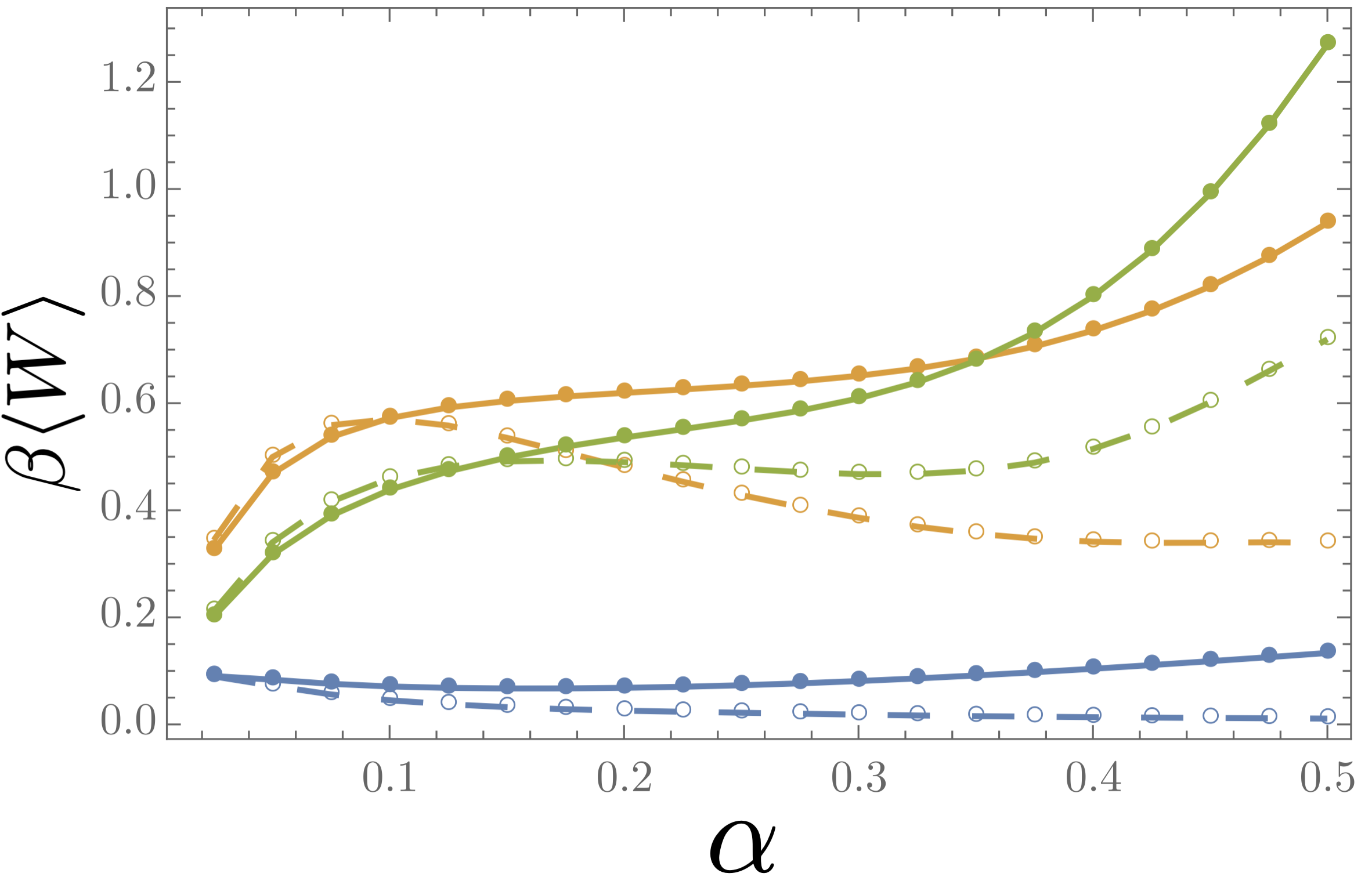}}
  \subfloat{\includegraphics[width=0.425\textwidth]{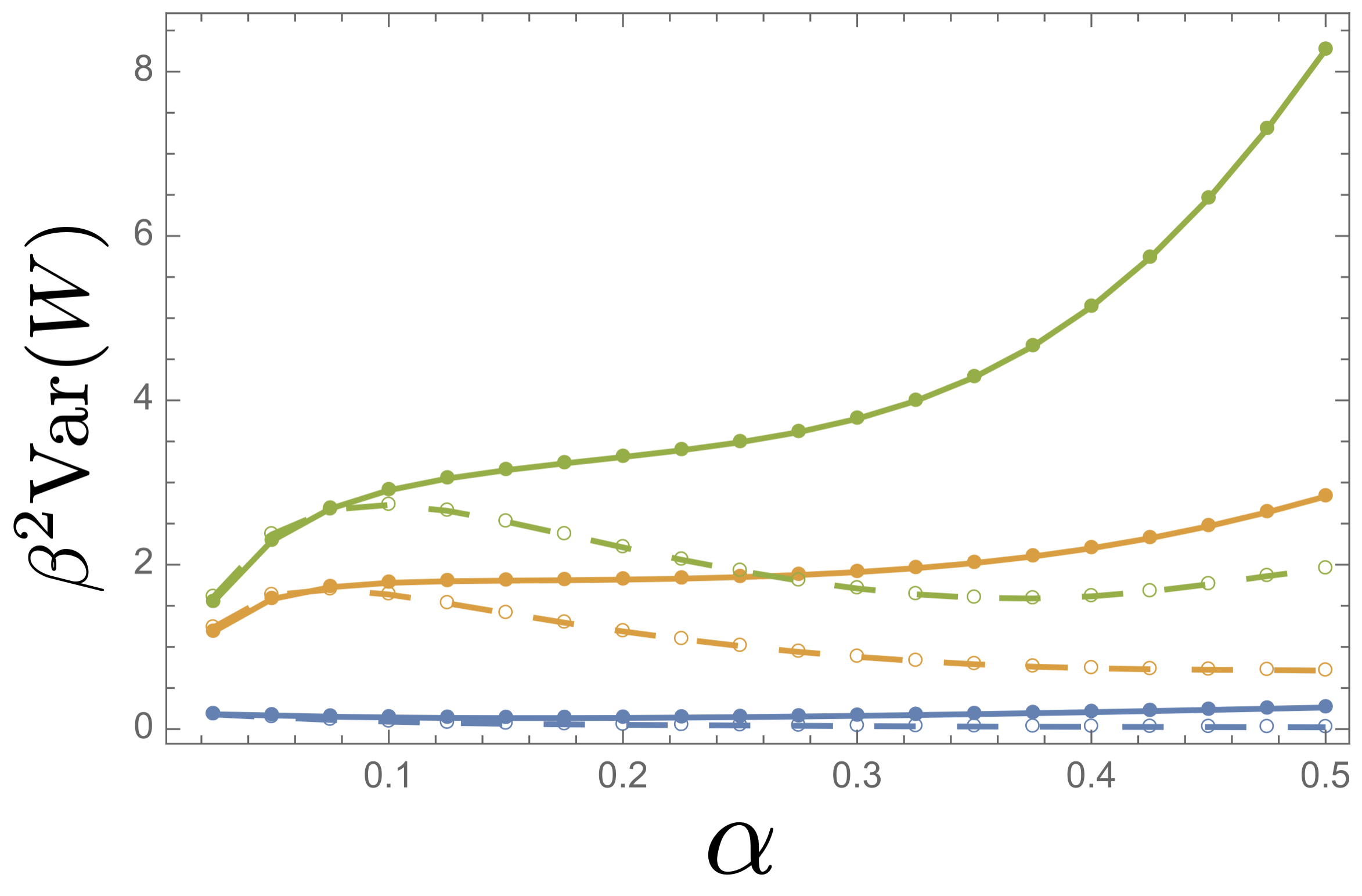}}
  \caption{Average work done per $k_B T$ (left) and work variance per $k_B^2 T^2$ (right) predicted by the polaron (solid line and solid markers) and weak-coupling (dashed line and open markers) generalised master equations as a function of the coupling strength $\alpha$. We plot results for the inverse temperature fixed at $\beta=1/(5\Delta)$ (blue), $\beta=1/(2\Delta)$ (orange), and $\beta=1/\Delta$ (green). Other parameters are as in Fig. 1 of the main text. \label{sup:grid}}
\end{figure}

\end{document}